\documentclass[conference]{IEEEtran}
\IEEEoverridecommandlockouts

\usepackage{cite}
\PassOptionsToPackage{numbers}{natbib}
\usepackage{amsmath,amssymb,amsfonts}
\usepackage{algorithmic}
\usepackage{graphicx}
\usepackage{textcomp}
\usepackage{xcolor}
\usepackage{hyperref}
\usepackage{soul}

\def\BibTeX{{\rm B\kern-.05em{\sc i\kern-.025em b}\kern-.08em
    T\kern-.1667em\lower.7ex\hbox{E}\kern-.125emX}}
\begin{document}

\title{QEEGNet: Quantum Machine Learning for Enhanced Electroencephalography Encoding \\
\thanks{This paper has been accepted by the 2024 IEEE Workshop on Signal Processing Systems (SiPS), 153-158. More details at \protect\url{https://ieeexplore.ieee.org/abstract/document/10768221/}.}
}

\author{
    \IEEEauthorblockN{Chi-Sheng Chen\IEEEauthorrefmark{1}\IEEEauthorrefmark{2}, Samuel Yen-Chi Chen\IEEEauthorrefmark{3},
     Aidan Hung-Wen Tsai\IEEEauthorrefmark{2}, 
     Chun-Shu Wei\IEEEauthorrefmark{1}}
    \IEEEauthorblockA{\IEEEauthorrefmark{1}Department of Computer Science, National Yang Ming Chiao Tung University, Hsinchu, Taiwan \\
    m50816m50816@gmail.com, wei@nycu.edu.tw}
    \IEEEauthorblockA{\IEEEauthorrefmark{3}Computational Science Initiative, Brookhaven National Laboratory, Upton NY, USA \\
    ycchen1989@ieee.org}
    
    \IEEEauthorblockA{\IEEEauthorrefmark{2}Neuro Industry, Inc., CA, USA \\
    \{michael, aidan\}@neuro-industry.com}
    
}

\maketitle

\begin{abstract}
Electroencephalography (EEG) is a critical tool in neuroscience and clinical practice for monitoring and analyzing brain activity. Traditional neural network models, such as EEGNet, have achieved considerable success in decoding EEG signals but often struggle with the complexity and high dimensionality of the data. Recent advances in quantum computing present new opportunities to enhance machine learning models through quantum machine learning (QML) techniques. In this paper, we introduce Quantum-EEGNet (QEEGNet), a novel hybrid neural network that integrates quantum computing with the classical EEGNet architecture to improve EEG encoding and analysis, as a forward-looking approach, acknowledging that the results might not always surpass traditional methods but it shows its potential. QEEGNet incorporates quantum layers within the neural network, allowing it to capture more intricate patterns in EEG data and potentially offering computational advantages. We evaluate QEEGNet on a benchmark EEG dataset, BCI Competition IV 2a, demonstrating that it consistently outperforms traditional EEGNet on most of the subjects and other robustness to noise. Our results highlight the significant potential of quantum-enhanced neural networks in EEG analysis, suggesting new directions for both research and practical applications in the field. 
\end{abstract}


\section{Introduction}
Electroencephalography (EEG) is a non-invasive technique widely used in neuroscience and clinical applications to measure electrical activity in the brain. The analysis of EEG data has been instrumental in understanding brain functions, diagnosing neurological disorders, and developing brain-computer interfaces. Traditional methods for EEG analysis often rely on conventional machine learning and deep learning techniques, such as the EEGNet model, which have demonstrated significant success in various EEG-based tasks. However, these models sometimes face limitations in capturing the complex and high-dimensional nature of EEG signals \cite{rashid_2020}.

Recent advancements in quantum computing have opened a new era for enhancing machine learning algorithms. Quantum machine learning (QML) leverages the principles of quantum mechanics to process information in fundamentally different ways compared to classical computing, offering potential advantages in terms of computational efficiency and the ability to explore larger solution spaces \cite{biamonte_2017}. The integration of QML with classical neural networks presents a promising hybrid approach that can potentially overcome some of the limitations of traditional deep learning models \cite{bharti2022noisy,cerezo2021variational}.

In this paper, we propose QEEGNet, a novel hybrid neural network that combines quantum machine learning techniques with the EEGNet architecture \cite{lawhern_solon_waytowich_gordon_hung_lance_2018} to enhance the encoding and analysis of EEG data. By incorporating quantum layers into the neural network, QEEGNet aims to leverage the power of quantum computing to improve the performance and robustness of EEG-based models. Our approach builds upon the strengths of EEGNet while introducing quantum elements that can capture more intricate patterns within EEG signals.

We evaluate QEEGNet on a famous benchmark EEG dataset, comparing its performance with the traditional EEGNet model. Our experimental results demonstrate that QEEGNet consistently outperforms these models in terms of accuracy and robustness to noise. These findings suggest that the integration of quantum machine learning can significantly enhance the capabilities of EEG analysis, paving the way for more effective and reliable applications in neuroscience and clinical settings.

The contributions of this paper are as follows:

\begin{itemize}
    \item We introduce Quantum-EEGNet (QEEGNet), a hybrid neural network that integrates variational quantum circuits (VQC) 
    with EEGNet for enhanced EEG encoding.
    \item  We provide a comprehensive evaluation of QEEGNet on a popular EEG open dataset, demonstrating its partial superior performance compared to traditional models.
    \item We discuss the practical feature embedding ability of all the models, prove that QEEGNet has the more advantage of feature representation than EEGNet in the field of EEG analysis.
\end{itemize}

\section{Related Work}

\subsection{Deep Learning on EEG Data}

The analysis of EEG data using neural networks has been a focal point in recent research \cite{li_chang_2019}, aiming to improve the accuracy and efficiency of EEG-based applications. Traditional models like artificial neural network, deep neural network models have shown significant promise but face limitations in handling the complex, high-dimensional nature of EEG signals. Recent advancements in quantum computing offer novel approaches to address these challenges.

EEGNet, tailored for classifying EEG Event-Related Potential (ERP) tasks, highlighting the potential of advanced neural networks in enhancing EEG signal processing. Their approach demonstrated improved classification accuracy, yet it also underscored the limitations of classical models in fully capturing the intricacies of EEG data.

There are several studies using deep learning dealing with different downstream tasks on EEG data. \cite{tosato_2023} using diffusion model to generate the EEG synthetic data, the proposed method could have a broader impact on neuroscience research by creating large, publicly available synthetic EEG datasets without privacy concerns. Artifact removal \cite{jiang_bian_tian_2019} is also a important field in EEG signal processing, \cite{lai_wang_yang_tsou_wei_2022} based on a pre-trained subject-independent model, was validated through multiple evaluations, showing it can maintain brain activity and outperform current artifact removal methods in decoding accuracy. In another study \cite{chen_wei_2024}, explored the integration of both graph attention and self-attention mechanisms with convolutional neural network (CNN) as EEG encoder for EEG-based image recognition. Their work showed that incorporating attention mechanisms can significantly enhance the model's ability to focus on relevant features of multimodal data, thereby improving performance. 

Despite these advancements, traditional deep learning models often fall short in fully addressing the high-dimensional and non-linear characteristics of EEG signals. As reviewed by \cite{khon_2023}, the incorporation of sophisticated deep learning components has shown potential benefits, yet there remains a need for more robust approaches to unlock the full potential of EEG data analysis. Recent advancements in quantum computing offer novel approaches to these challenges. Quantum machine learning (QML) models, with their ability to handle high-dimensional data and complex dependencies, present a promising direction for advancing EEG data analysis. Notably, \cite{anupamapadha_sahoo_2024} demonstrates the potential of QML approaches in time-series signal processing across several models, further underscoring the capabilities of QML in addressing the limitations of classical neural networks. By leveraging the principles of quantum mechanics, QML models can potentially overcome the limitations of classical neural networks, paving the way for more accurate and efficient EEG-based applications.
%
\subsection{Quantum Machine Learning on EEG Data}
Quantum machine learning (QML) represents a cutting-edge advancement that can further elevate EEG analysis. QML algorithms, especially the hybrid quantum-classical algorithms based on VQC, draw inspiration from several traditional deep learning algorithms, such as quantum convolutional neural networks (QCNN) \cite{cong_choi_lukin_2019}, quantum generative adversarial networks (QGAN) \cite{zoufal_lucchi_woerner_2019}, and quantum long short-term memory networks (QLSTM) \cite{samuelyen-chen_2022}, among others.
%
A recent paper \cite{mari_bromley_izaac_schuld_killoran_2020} discussed the application of hybrid quantum-classical neural networks for pattern recognition tasks. In this setting, a classically pre-trained model is used to preprocess the data, which is then sent to a variational quantum circuit (VQC) to further process. Their findings suggest that incorporating quantum layers can enhance classical models by providing greater computational power and the ability to explore larger solution spaces. Applying this hybrid quantum-classical approach to EEG analysis can potentially address the intricate patterns and dependencies in EEG data, potentially overcoming the limitations faced by traditional neural networks.

Moreover, QML has shown promise in improving model classification performance. 
\cite{yousif_belal_2024} provided a QCNN to do the medical image classification. In the EEG QML application, \cite{toshiaki_2022} proposed a hybrid quantum-classical neural networks with a VQC in front of the traditional neural network, \cite{ho_hung_2023} and \cite{aksoy_karabatak_2024} exploring EEG classification task using quantum support vector machine (QSVM), \cite{lins_mendes_bezerra_ramos_das_2024} using different quantum circuit architectures as feature extractor with the classic multilayer perceptron (MLP) to do drowsiness detection. These researches emphasizing the potential to revolutionize various machine learning tasks, including those involving complex, high-dimensional data such as EEG. Their findings suggest that quantum algorithms can significantly enhance the representational power and efficiency of traditional neural networks, paving the way for more robust EEG analysis model. 
These studies collectively highlight the evolving landscape of EEG analysis, where integrating advanced neural network architectures and quantum computing techniques can lead to substantial performance gains. Building upon these foundations, our work differentiates from these aforementioned studies by specifically integrating VQC within the EEGNet architecture to create QEEGNet. While previous research has shown the promise of QML in various domains, including time series analysis and medical imaging, our approach focuses on leveraging quantum layers to enhance EEG encoding and analysis. By doing so, we aim to address the unique challenges posed by EEG data, such as its high dimensionality and complex temporal-spatial dependencies.
The novelty of QEEGNet lies in being the first quantum-classical hybrid model that incorporates VQC quantum encoding layers at the end of the model. This unique architecture enables QEEGNet to capture more intricate patterns within EEG signals, improving performance and robustness in EEG-based applications. Our experimental results demonstrate that QEEGNet consistently outperforms traditional EEGNet models, highlighting the significant potential of quantum-enhanced neural networks in the field of EEG analysis. 

\section{Methodology}
\begin{figure}
    \centering
    \includegraphics[width=1\linewidth]{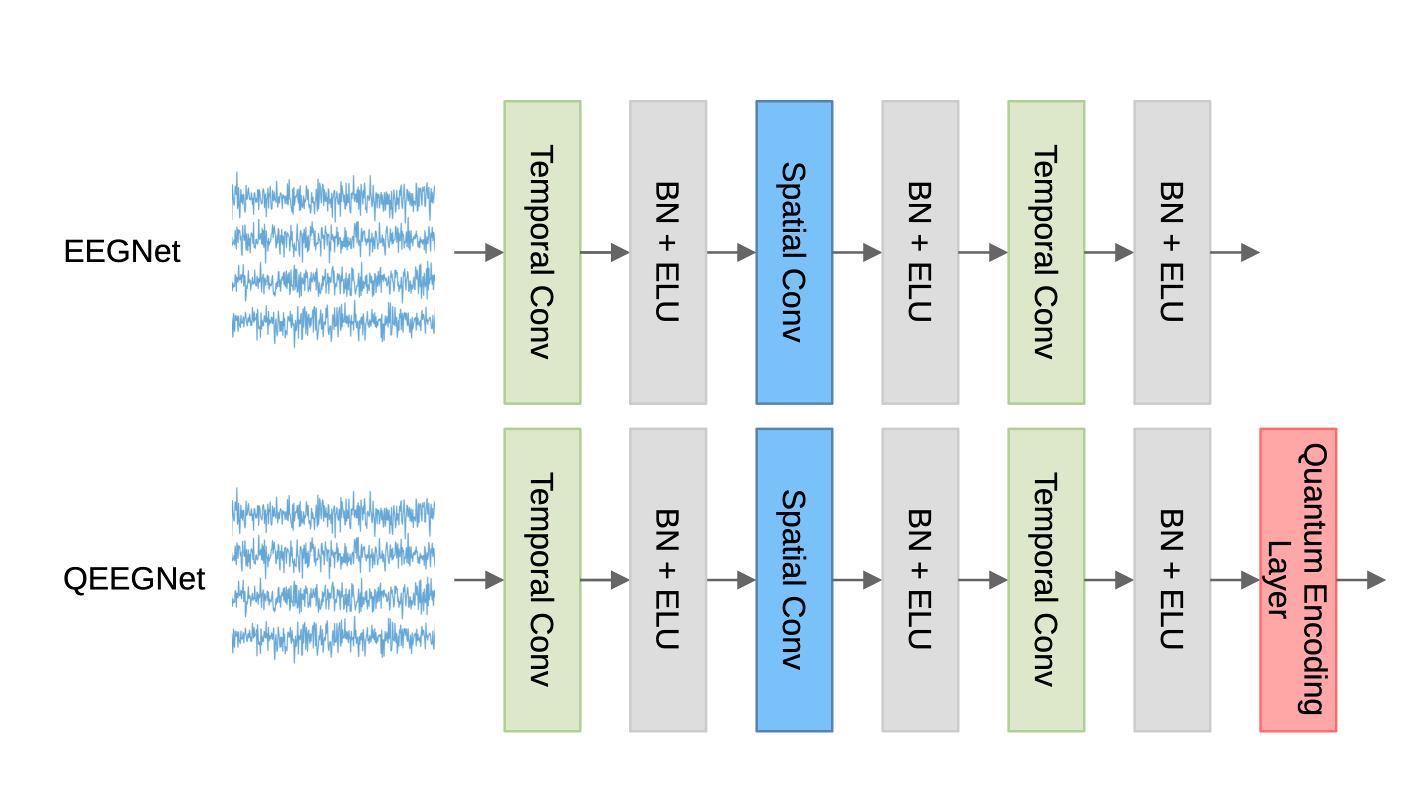}
    \caption{Schematic illustration comparison of the EEGNet and the proposed Quantum-EEGNet (QEEGNet). In this diagram, Conv denotes convolution, BN denotes batch normalization, and ELU denotes the exponential linear unit activation function.}
    \label{fig:eegnet_qeegnet}
\end{figure}
In this section, we detail the architecture and methodology of QEEGNet, our proposed hybrid quantum-classical neural network model that leverages quantum machine learning to enhance EEG encoding and analysis.

\subsection{Architecture of EEGNet}\label{AA}
In the field of brain-computer interface (BCI) and EEG signal processing, EEGNet \cite{lawhern_solon_waytowich_gordon_hung_lance_2018} is a popular method, it a compact CNN model specifically designed for the analysis and classification of EEG data. The network consists of only a few layers, each designed to capture different aspects of the EEG signal, including temporal and spatial features. The EEGNet architecture illustrated in Fig~\ref{fig:eegnet_qeegnet}.
\subsection{Quantum Circuit}
Quantum circuits can potentially provide computational advantages over classical methods for specific tasks. By exploiting the superposition and entanglement of qubits, quantum circuit can represent and process information in ways that classical bits cannot. This capability is mathematically represented as:
\begin{equation}
\text{Quantum State: } |\psi\rangle = \alpha|0\rangle + \beta|1\rangle,
\end{equation}
where $\alpha$ and $\beta$ are complex amplitudes, and $|\psi\rangle$ can exist in multiple states simultaneously.
Quantum entanglement and superposition enable the exploration of complex data structures and correlations. For instance, a quantum state with $n$ qubits can represent $2^n$ states simultaneously, providing a richer feature space for learning tasks:
\begin{equation}
|\psi\rangle = \sum_{i=0}^{2^n-1} \alpha_i |i\rangle,
\end{equation}
where \( |i\rangle \) represents the basis states of the quantum system, and \( \alpha_i \) are the coefficients that determine the probability amplitudes of these states in the superposition. The index \( i \) helps to enumerate all possible states of the \( n \)-qubit system, which can be in any combination of the \( 2^n \) possible states. 

This superposition allows the quantum encoding layer to encode and process more information than a classical layer of comparable size.
\begin{figure*}
    \centering
    \includegraphics[width=1\linewidth]{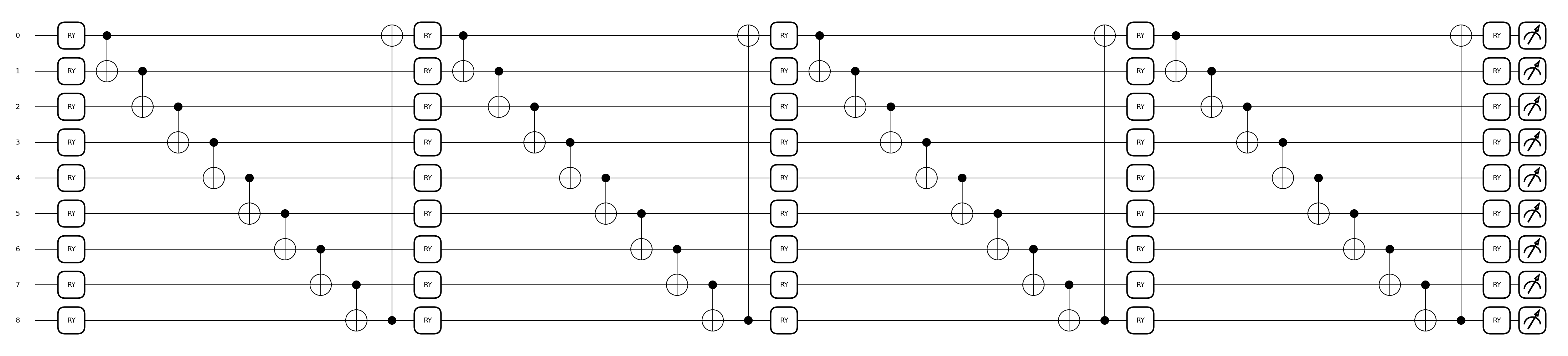}
    \caption{The illustration of quantum encoding layer module of the proposed QEEGNET. The module contains nine qubits and four quantum circuit layers in this work.}
    \label{fig:qlayer_cir}
\vskip -0.1in
\end{figure*}
\subsection{Quantum Encoding Layer Module}
The quantum encoding layer module shows in in Fig~\ref{fig:qlayer_cir}. It is to integrate quantum computing capabilities into classical neural networks. By incorporating a quantum circuit within a classical deep learning model, the quantum encoding layer aims to leverage the unique properties of quantum mechanics, such as superposition and entanglement, to enhance the performance and capabilities of machine learning models. It leverages a parameterized quantum circuit defined by mapping classical EEG features as a set of qubits $\mathbf{x} \in \mathbb{R}^{n_{\text{qubits}}}$ and has trainable weights $\mathbf{w} \in \mathbb{R}^{n_{\text{layers}} \times n_{\text{qubits}}}$. The first step in the quantum encoding layer is encoding the input. Each qubit \( q_i \) is rotated based on the input data \( x_i \) using the rotation gate \( \text{RY} \) (rotation-Y gate):
\begin{equation}
    RY(x_i) = \exp\left(-i \frac{x_i}{2} \sigma_y\right)
,\end{equation}
    where \( \sigma_y \) is the Pauli-Y matrix:
\begin{equation}
    \sigma_y = \begin{pmatrix}
    0 & -i \\
    i & 0
    \end{pmatrix}.
\end{equation}
Second, there are some parameterized layers in the quantum encoding layer module, in this QEEGNet model, we use a ring pattern of 
CNOT (controlled-NOT) gates. The CNOT gate is a two-qubit gate where the state of one qubit (the control qubit) determines whether to flip the state of another qubit (the target qubit). The CNOT gate is a two-qubit quantum gate represented by the following 4x4 unitary matrix:
\begin{equation}
{CNOT} = \begin{pmatrix}
1 & 0 & 0 & 0 \\
0 & 1 & 0 & 0 \\
0 & 0 & 0 & 1 \\
0 & 0 & 1 & 0
\end{pmatrix}.
\end{equation}
A ring pattern involves applying CNOT gates such that each qubit is entangled with its neighbor, and the last qubit is entangled with the first qubit, forming a closed loop or "ring." This pattern ensures that entanglement is spread throughout the entire set of qubits. For each layer \( l \) in \( n_{\text{layers}} \), a ring pattern of \(\text{CNOT}\) gates entangles the qubits:
\begin{equation}
      CNOT(q_i, q_{(i+1) \mod n_{\text{qubits}}}) \quad \text{for} \quad i = 0, 1, \ldots, n_{\text{qubits}}-1.
\end{equation}
Each qubit $q_i$ undergoes an additional rotation based on the trainable weight $w_{l,i}$:
\begin{equation}
RY(w_{l,i}) = \exp\left(-i \frac{w_{l,i}}{2} \sigma_y\right).
\end{equation}
The final stage of the quantum encoding layer module is measurement the states of the qubits. The expectation value of the Pauli-Z operator $\sigma_z$ is measured for each qubit, yielding the output:
\begin{equation}
\langle \sigma_z^i \rangle = \langle 0 | U^\dagger \sigma_z^i U | 0 \rangle,
\end{equation}
where $U$ is the unitary operation representing the quantum circuit.
The Pauli-Z operator, also known as the Pauli-Z matrix, is represented by the following 2x2 matrix:
\begin{equation}
\sigma_z = \begin{pmatrix}
1 & 0 \\
0 & -1
\end{pmatrix}.
\end{equation}
This operator is used to measure the z-component of the spin of a qubit and has eigenvalues of +1 and -1, corresponding to the basis states \(|0\rangle\) and \(|1\rangle\), respectively. 

\subsection{Architecture of Quantum-EEGNet}
The integration of quantum circuits into classical neural networks enables the creation of hybrid models that combine the strengths of both paradigms. We provided a novel classical-to-quantum Data encoding on EEG signal processing. The quantum encoding layer serves as a bridge, allowing the network to learn quantum-encoded features while being trained using classical optimization techniques. The overall architecture of QEEGNet is shown in TABLE~\ref{tab:qeegnet}. The input EEG data is first encoded by the classical EEGNet. The output features are then encoded into qubits through the quantum encoding layer module, leveraging the high-dimensional information inherent in Hilbert space. The last stage of the model is output the quantum measurement result into fully connected layer to do the downstream classification task.

\begin{table}[ht]
\caption{Network structure of QuantumEEGNet}
\centering
\scalebox{0.87}{
\begin{tabular}{|c|c|c|c|c|}
\hline
Layer & Input $(C \times T)$ & Operation & Output  \\
\hline
1 & $1 \times 1 \times T$ & Conv2D (1, 64) & $16 \times 1 \times (T-63)$ \\
  & $16 \times 1 \times (T-63)$ & BatchNorm & $16 \times 1 \times (T-63)$ \\
  & $16 \times 1 \times (T-63)$ & ELU & $16 \times 1 \times (T-63)$  \\
\hline
2 & $16 \times 1 \times (T-63)$ & ZeroPad2D (16, 17, 0, 1) & $16 \times 1 \times (T-30)$ \\
  & $16 \times 1 \times (T-30)$ & Conv2D (2, 32) & $32 \times 1 \times (T-61)$  \\
  & $32 \times 1 \times (T-61)$ & BatchNorm & $32 \times 1 \times (T-61)$ \\
  & $32 \times 1 \times (T-61)$ & ELU & $32 \times 1 \times (T-61)$  \\
  & $32 \times 1 \times (T-61)$ & MaxPool2D (2, 4) & $32 \times 1 \times \frac{(T-61)}{4}$  \\
  & $32 \times 1 \times \frac{(T-61)}{4}$ & Dropout (0.25) & $32 \times 1 \times \frac{(T-61)}{4}$  \\
\hline
3 & $32 \times 1 \times \frac{(T-61)}{4}$ & ZeroPad2D (2, 1, 4, 3) & $32 \times 1 \times \frac{(T-52)}{4}$  \\
  & $32 \times 1 \times \frac{(T-52)}{4}$ & Conv2D (8, 4) & $32 \times 1 \times \frac{(T-55)}{4}$  \\
  & $32 \times 1 \times \frac{(T-55)}{4}$ & BatchNorm & $32 \times 1 \times \frac{(T-55)}{4}$ \\
  & $32 \times 1 \times \frac{(T-55)}{4}$ & ELU & $32 \times 1 \times \frac{(T-55)}{4}$  \\
  & $32 \times 1 \times \frac{(T-55)}{4}$ & MaxPool2D (2, 4) & $32 \times 1 \times \frac{(T-55)}{16}$  \\
  & $32 \times 1 \times \frac{(T-55)}{16}$ & Dropout (0.25) & $32 \times 1 \times \frac{(T-55)}{16}$  \\
\hline
4 & $32 \times 1 \times \frac{(T-55)}{16}$ & Quantum Encoding Layer & $9 \times N$ \\
\hline
5 & $9 \times N$ & Fully Connected Layer & $N$  \\
\hline
\end{tabular}
}
\label{tab:qeegnet}
\end{table}

\section{Results and Discussions}
\begin{table*}[t]
    \caption{Comparison of the highest validation accuracy results of EEGNet and QEEGNet on the BCIC-IV-2a Dataset.}
    \centering
    \begin{tabular}{ccccccccccc}
                &  Subject 1 & Subject 2 & Subject 3 & Subject 4 & Subject 5 & Subject 6 & Subject 7 & Subject 8 & Subject 9 & Average\\
       EEGNet   &  42.7\% & 25.0\% & 46.4\% & 32.1\% & 25.0\% & \textbf{42.9\%} & \textbf{42.9\%} & \textbf{32.1\%} & 69.6\% & 39.8\% \\
       QEEGNet  & \textbf{50.0\%} & \textbf{26.8\%} & \textbf{50.0\%} & \textbf{35.7\%} & \textbf{32.1\%} & 41.1\% & 39.3\%  & 30.4\% & \textbf{73.2\%} & \textbf{42.1\%}\\
    \end{tabular}
    \label{tab:qeegnet_val_res}
\end{table*}

\begin{table*}[t]
    \caption{Comparison of the highest test accuracy results of EEGNet and QEEGNet on the BCIC-IV-2a Dataset.}
    \centering
    \begin{tabular}{ccccccccccc}
                &  Subject 1 & Subject 2 & Subject 3 & Subject 4 & Subject 5 & Subject 6 & Subject 7 & Subject 8 & Subject 9 & Average\\
       EEGNet  &  47.9\%  & 22.9\% & \textbf{46.5\%} & \textbf{32.6\%} & 26.4\% & 28.8\% & 33.7\% & \textbf{32.3\%} & 62.2\% & 37.7\%\\
       QEEGNet &  \textbf{49.3\%} & \textbf{30.2\%} & 44.1\% & 30.6\% & \textbf{26.7\%} & \textbf{31.9\%} & \textbf{36.8\%} & 28.1\% & \textbf{65.3\%} &  \textbf{38.1\%}\\
    \end{tabular}
    \label{tab:qeegnet_test_res}
\end{table*}
\subsection{BCIC-IV-2a Dataset}
We use the BCIC-IV-2a (Brain-Computer Interface Competition) dataset, which provides time-asynchronous EEG data. This dataset is one of the most popular public EEG datasets, released for the BCI Competition IV in 2008 \cite{tangermann_2012}. It includes EEG recordings from nine subjects who performed a four-class motor-imagery task, repeated twice on different days. During the task, subjects imagined one of four movements (right hand, left hand, feet, and tongue) for four seconds after a cue. Each session had 288 trials, with 72 trials for each movement type. The EEG signals were recorded using 22 electrodes placed around the central region at a sampling rate of 250 Hz. We processed the EEG signals by down-sampling from 250 Hz to 128 Hz, applying a band-pass filter at 4-38 Hz, and segmenting the signals from 0.5 to 4 seconds after the cue, resulting in 438 time points per trial. 

\subsection{Experiment Details}
For BCIC-IV-2a dataset, we used the first session of a subject for the training set, with one-fifth of it set aside for validation. Using one-fifth for validation instead of the more common one-eighth or one-ninth seen in some documents is intended to increase the difficulty of training and to increase the number of validation samples, with the hope of better highlighting the differences and generalization between quantum and classical models. The model that had the lowest validation loss within 100 epochs was then tested on the second session of the same subject. We trained QEEGNet, which has 9 qubits and 4 quantum layers. All the models are  implemented by Pytorch and PennyLane frameworks with the simulator as backend. Input batch size for training is 32 with training 100 epochs, using AdamW as optimizer with $10^{-3}$ learning rate. We selected the model with the highest validation accuracy for predicting on the test dataset. The training time for QEEGNet per subject was approximately 20 hours on CPUs, using a Google Cloud Platform a2-ultragpu-1g machine with 170 GB of RAM.

\subsection{Results}
The experiment results are shown in TABLE~\ref{tab:qeegnet_val_res} and TABLE~\ref{tab:qeegnet_test_res}. The results highlight QEEGNet's consistent performance advantages over EEGNet in both validation and test datasets. This performance disparity can be attributed to the quantum layer integrated into QEEGNet, which likely enhances its ability to capture complex patterns and features within EEG data. The substantial improvements in validation and test accuracies for multiple subjects suggest that QEEGNet can generalize better across different data variations. Moreover, the noticeable improvement in subjects with lower accuracies using EEGNet (e.g., Subject 2) emphasizes QEEGNet's robustness and potential for broader applicability. 

\subsection{Model Interpretation and Discussion}
\begin{figure}
    \centering
    \includegraphics[width=.8\linewidth]{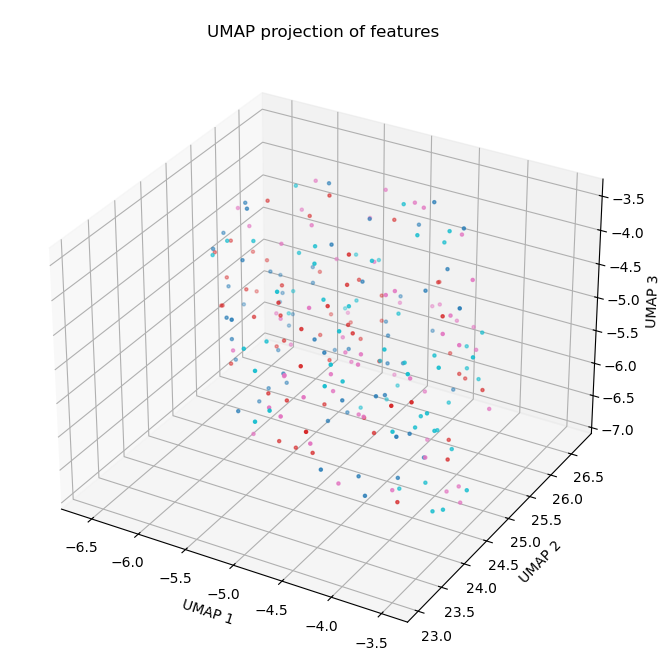}
    \caption{3D-UMAP projection of EEGNet features.  Different classes are marked by different colors.}
    \label{fig:cls_umap}
\end{figure}
\begin{figure}
    \centering
    \includegraphics[width=.8\linewidth]{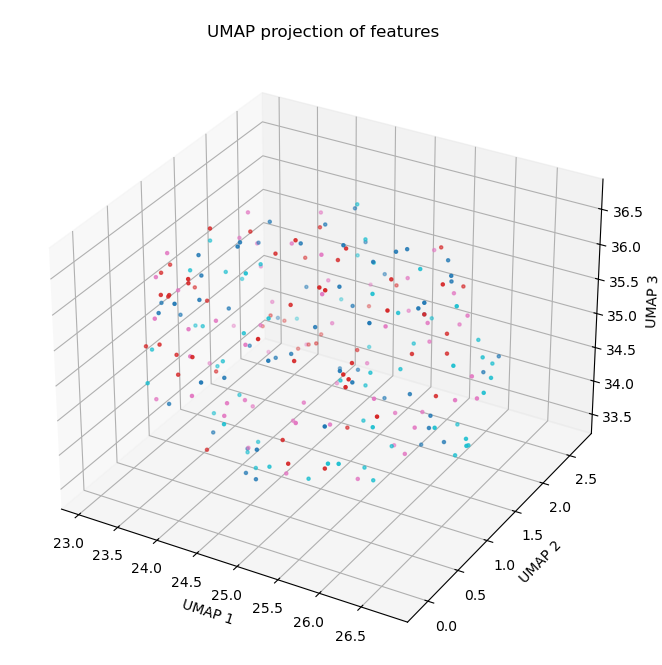}
    \caption{3D-UMAP projection of QEEGNet features. Different classes are marked by different colors.}
    \label{fig:qeegnet_umap}
\end{figure}

We use uniform manifold approximation and projection (UMAP) \cite{mcinnes_healy_melville_2018} method to compare the model embedding performance between two models. The provided UMAP projections visually represent the feature embeddings produced by EEGNet and QEEGNet models. These 3D scatter plots help in understanding the underlying structure and separability of the data as transformed by each model.
The first UMAP projection in Fig~\ref{fig:cls_umap}. showcases the embeddings generated by the EEGNet model. The points in this plot are distributed with a noticeable spread, indicating that EEGNet captures a variety of features from the input data. However, the embeddings exhibit some degree of overlap, suggesting potential challenges in distinguishing between different classes or patterns. The spread of points suggests a moderate clustering tendency, but with some intermingling between clusters, indicating that while EEGNet can learn useful features, its ability to distinctly separate different classes might be limited.
In contrast, the second UMAP projection in Fig~\ref{fig:qeegnet_umap}. depicts the embeddings from the QEEGNet model. The points in this plot appear more tightly clustered compared to the EEGNet embeddings. This suggests that QEEGNet has a better capability to group similar features together, enhancing the separability between different classes or patterns. The clusters are more distinct and less overlapping, indicating that QEEGNet is more effective in capturing and distinguishing between complex features in the data.
The comparison between the two UMAP projections highlights the improvements brought by integrating quantum layers into the QEEGNet model. QEEGNet's embeddings show a more pronounced clustering effect, which can be attributed to its enhanced feature extraction capabilities. The tighter and more distinct clusters suggest that QEEGNet is better at capturing the underlying structure of the EEG data, leading to improved performance in classification tasks, as evidenced by the higher validation and test accuracies discussed previously. Furthermore, the clearer separation in QEEGNet's embeddings implies that the model can more effectively learn and represent the unique characteristics of different classes. This can lead to more robust and reliable predictions, particularly in complex datasets where traditional models like EEGNet might struggle.

\section{Conclusion}
In this paper, we introduced QEEGNet, a novel hybrid neural network that integrates quantum computing with the traditional EEGNet architecture to enhance the encoding and analysis of EEG data. Our experimental results on the BCIC-IV-2a dataset demonstrate that QEEGNet overall outperforms traditional EEGNet in terms of accuracy and robustness across most subjects. The integration of quantum layers within the neural network allows QEEGNet to capture more intricate patterns in EEG data, suggesting a significant potential for quantum-enhanced neural networks in the field of EEG analysis.

The findings highlight the practical feature embedding ability of QEEGNet, showcasing its advantage in feature representation over traditional models. The UMAP projections further validate these improvements by illustrating the superior clustering and separability of features learned by QEEGNet. These enhancements underline the potential of quantum machine learning to provide computational advantages and explore larger solution spaces, paving the way for more effective and reliable applications in neuroscience and clinical settings.

The performance improvements achieved by QEEGNet suggest new directions for research and practical applications, emphasizing the importance of further exploring and developing quantum-enhanced models for complex and high-dimensional data analysis tasks. The study opens up opportunities for integrating advanced quantum computing techniques with classical neural networks to achieve substantial gains in performance and robustness in EEG signal processing and beyond.

\bibliographystyle{IEEEtran}
\bibliography{references,qml}

\end{document}